\documentclass{pasa}%

\usepackage{graphicx,color}

\title[The Core-Cusp Problem Revisited: ULDM vs. CDM]{The Core-Cusp Problem Revisited: ULDM vs. CDM}

\author[Emily Kendall and Richard Easther]{Emily Kendall$^1$, Richard Easther$^1$, \affil{$^1$Department of Physics, University of Auckland, Private Bag 92019, Auckland, New Zealand}}%

\jid{PASA}
\doi{10.1017/pas.\the\year.xxx}
\jyear{\the\year}

\usepackage{aas_macros}
\usepackage{hyperref} 
\hypersetup{colorlinks,citecolor=blue,linkcolor=blue,urlcolor=blue}

\hypersetup{draft}
\setcitestyle{numbers,open={[},close={]}}

\begin{document}

\begin{frontmatter}
\maketitle

\begin{abstract}
The core-cusp problem is a widely cited motivation for the exploration of dark matter models beyond standard CDM. One such alternative is ULDM; extremely light scalar particles exhibiting wavelike properties on kiloparsec scales.  Astrophysically realistic ULDM halos are expected to consist of inner solitonic cores embedded in NFW-like outer halos. The presence of the solitonic core suggests that ULDM may resolve the core-cusp discrepancy associated with pure NFW halos without recourse to baryonic physics. However, it has been demonstrated that the density of ULDM halos can exceed those of comparable NFW configurations at some radii and halo masses, apparently exacerbating the problem rather than solving it. This situation arises because, although solitonic cores are flat at their centres, they obey an inverse mass-radius scaling relationship. Meanwhile, the mass of the inner soliton increases with the total halo mass, and therefore the inner core becomes more peaked at large halo masses. We describe a parameterisation of the radial density profiles of ULDM halos that allows for environmental variability of the core-halo mass relation in order to investigate this issue in more detail.  For halos up to $10^{12} M_\odot$ we find feasible ULDM profiles for which the central density is lower than their NFW counterparts at astrophysically accessible radii. However, comparisons to observed profiles do not strongly favour either option; both give reasonable fits to subsets of the data for some parameter choices. Consequently, we find that robust tests of the core-cusp problem in ULDM will require more comprehensive observational data and simulations that include baryonic feedback. 

\end{abstract}

\end{frontmatter}

\section{INTRODUCTION }
\label{sec:intro}

It is widely agreed that non-baryonic dark matter constitutes the majority of the mass of the observable universe, but its precise nature remains an open question. Many dark matter models have been proposed, with particle CDM [Cold Dark Matter]  being the most widely studied. This scenario successfully accounts for the large scale structure of the universe \cite{Springel:2005nw} and the spectrum of anisotropies in the microwave background \cite{deBernardis:2000sbo, Hanany:2000qf, Halverson:2001yy, Netterfield:2001yq, Lee:2001yp, Ade:2015xua,  Hu:2001bc}. Nevertheless, the so-called ``small-scale crisis'' remains a challenge \cite{Weinberg:2013aya}. A key issue is the apparent tension between the central density profiles of dark matter halos in simulations containing only gravitationally interacting CDM, and those inferred from observational data. Simulations tend to produce `cuspy' central density profiles \cite{Navarro:1995iw}, which grow as $1/r$ at small radii, but observational data appears to favour flattened central cores \cite{Moore:1994yx}. The tension between the two is widely known as the ``core-cusp problem'' \cite{Dutton:2018nop, Read:2018pft, Genina:2018}. 
 
The seriousness of the core-cusp problem is the subject of  ongoing debate, as it has been shown that it may be ameliorated in some cases by adding baryonic matter to CDM simulations \cite{Benitez-Llambay:2018}. Nevertheless, the wider category of ``small-scale'' problems in standard CDM, along with tighter constraints from direct-detection experiments \cite{Schumann:2019eaa}, motivates the study of alternative dark matter models. One scenario which has gained substantial traction is ultra-light dark matter [ULDM], also known as scalar-field dark matter, $\Psi$ dark matter, BEC dark matter and fuzzy dark matter. Often these models are referred to as `axion-like', given that they describe an extremely light scalar field. %We adopt the the terminology `ULDM', emphasisin the extremely small mass of the constituent particles, but does not necessarily conflate these with the QCD axion. 

As reviewed by Hui {\em et al.\/} \cite{Hui:2016ltb}, current constraints prefer a ULDM particle mass of $\mathcal{O}(\sim 10^{-22}\mathrm{eV})$, corresponding to a kiloparsec-scale de Broglie wavelength. ULDM thus exhibits novel wave-like behaviour on astrophysically interesting scales, as well as supporting soliton-like gravitationally confined Bose-Einstein condensates. ULDM simulations suggest that realistic astrophysical halos have an inner core consisting of a kiloparsec scale condensate, surrounded by a virialised outer halo \cite{Schwabe:2016rze, Veltmaat:2018dfz}. The outer region resembles a standard CDM halo which is well approximated by the NFW profile characteristic of collisionless CDM, and most commonly associated with WIMP dark matter [Weakly Interacting Dark Matter] \cite{Navarro:1995iw}. 
Because the (solitonic) profiles of the inner condensates are flat, it has been suggested that ULDM can resolve the core-cusp problem without the need to invoke baryonic astrophysics. However, solitonic density profiles obey an inverse mass-radius scaling law, so the density of the ULDM halo might exceed that of an analogous NFW halo over a finite range of small radii in larger galaxies. In particular, Ref~\cite{Robles:2018fur} concludes that DM-only NFW profiles may outperform ULDM profiles for galaxies with halo masses $\mathrm{M_h} \gtrsim 10^{11} \mathrm{M}_{\odot}$.\footnote{ Ref~\cite{Robles:2018fur}  refers to these galaxies as ``large dwarfs'', though we note that the upper limit on this category is around $10^{12} \mathrm{M_{\odot}}$, approaching estimates of the Milky Way mass \cite{Watkins2019ApJ}.}

To further explore the possible worsening of the core-cusp problem in ULDM, we examine the effect of scatter in the core-halo mass scaling relation. Starting from the semi-analytic density profile of Ref.~\cite{Robles:2018fur}, we look at the scatter in the parameters implied by Ref.~\cite{Schive:2014hza}. We show that the resulting statistical variability may ease concerns that the core-cusp problem is exacerbated for ULDM relative to CDM for ``large dwarf'' galaxies. 

Our analysis also highlights a number of caveats that apply to all such discussions. First, the incoherent outer regions of ULDM halos are subject to strong fluctuations, both temporally and spatially. These are not captured by semi-analytic halo density profiles and we argue that these fluctuations may accentuate the intrinsic scatter in halo parameters. Moreover, baryonic feedback is known to be significant for dwarf galaxies \cite{2018MNRAS.473.5698D, Benitez-Llambay:2018} and neither the NFW or ULDM profiles incorporate this effect. Consequently, we caution against attempting to discriminate between ULDM and CDM models based on DM-only simplified theoretical profiles. 

Observationally, we find that neither semi-analytic ULDM halos (for ULDM particle mass $0.8-2.5\times 10^{-22} \operatorname{eV}$) nor NFW halos provide a particularly convincing fit to rotation curves of large dwarf galaxies in the SPARC database \cite{Lelli:2016zqa}. Moreover, many rotation curves are extracted from a few data points with significant uncertainties and which only span a small range of radial distances, further complicating attempts to draw robust conclusions. These issues are exacerbated by the relatively large number of free parameters in the theoretical models. For instance, a ULDM mass parameter of $10^{-23}\operatorname{eV}$ seemingly ameliorates the core-cusp problem in galaxies exhibiting a steep decrease in rotation velocity at small radii, but such a small mass is in tension with other constraints. Consequently, the primary conclusion to be drawn from this type of parameter-fitting exercise seems to be that  analyses of the core-cusp problem (and potentially other ``small scale'' anomalies) based on simplified semi-analytic  DM-only models cannot meaningfully test these scenarios, especially when observational data is limited and detailed numerical simulations with baryonic feedback have not been performed. 

The structure of the paper is as follows. In Section \ref{sec:models}, we review the construction of semi-analytic density profiles for both the ULDM and CDM models and briefly discuss aspects of realistic ULDM halos which are not captured by the semi-analytic model. In Section \ref{sec:velocity} we compare the semi-analytic density profiles for ULDM and CDM halos in the dwarf galaxy mass range $10^{11} - 10^{12}\operatorname{M}_{\odot}$, taking into account statistical variation in both the NFW concentration parameter and the ULDM core-halo mass relation. We then compare the radial velocity profiles inferred from these density profiles with astrophysical data from the SPARC database \cite{Lelli:2016zqa}. We conclude in Section \ref{sec:conclusion}.

\section{Semi-analytic Halos}\label{sec:models}

%\subsection{The NFW profile of CDM}\label{sec:NFW}

We begin by looking at the semi-analytic parametrisations of ULDM and CDM halos. The  well known  NFW   profile of CDM \cite{Navarro:1995iw, Maccio:2008pcd} is given by
\begin{equation}\label{eq:nfw}
    \rho_\mathrm{NFW}(r)=\frac{\rho_0}{\frac{r}{R_s}\left(1+\frac{r}{R_s}\right)^2} \, .
\end{equation}
At small radii the profile is proportional to $1/r$, while at large radii it goes as $1/r^3$. The parameters $\rho_0$ and $R_s$ vary from halo to halo; $\rho_0$ can be interpreted as a characteristic density, while $R_s$ is the scale radius and determines the distance from the centre at which the transition between the `small $r$' and `large $r$' limits occurs. 

The NFW halo is assumed to be radially symmetric, and requires truncation at a finite radius in order to prevent the integrated mass diverging as $r\rightarrow \infty$. The truncation is typically set by the virial radius, which is itself determined approximately via the spherical top-hat collapse model describing the evolution of a uniform spherical overdensity in a smooth expanding background \cite{White:2000jv, Suto:2015jdt, Herrera:2017epn}. Gravitational collapse of the overdensity halts when virial equilibrium is reached. In this scenario the corresponding virial radius is the radius at which the mean internal density is $\Delta_c \rho_\mathrm{crit}(t)$. Here $\rho_\mathrm{crit}(t)$ is the critical density of the universe at time $t$. The  factor $\Delta_c$ is of order $10^2$ and while different conventions exist, we  make the common choice $\Delta_c = 200$ \cite{Richings:2018} in what follows. 

Once the virial radius is specified as the outer limit of the halo, Equation \ref{eq:nfw} completely determines the density profile for given $\rho_0$ and $R_s$. For any given virial mass, there is a range of corresponding NFW density profiles, with the distributions of $\rho_0$ and $R_s$ emerging from the mass-concentration-redshift relation seen in N-body simulations and observations \cite{Ludlow:2013vxa, Ragagnin:2018enf}. 

%\subsection{The piecewise ULDM halo profile}

Whereas CDM halos can be described by NFW distributions, a different approach must be taken in the case of ULDM. ULDM dynamics is governed by the Schr{\"o}dinger-Poisson system of coupled differential equations. In a static background, they take the dimensionless form  
\begin{align}
    &i\dot{\psi} = -\frac{1}{2}\nabla^2\psi+\Phi\psi,\\
    &\nabla^2\Phi = 4\pi \vert \psi\vert^2,
\end{align}
where $\psi$ is the ULDM wavefunction, $\Phi$ is the Newtonian potential, and the density $\rho \propto |\psi|^2$. The solitonic ground state profile cannot be written down analytically, but given a numerically computed spherically symmetric profile $\psi$ for $\psi(0)=1$, the full family of solutions is then given by
\begin{equation}
    \psi'(x) = \gamma\psi(\sqrt{\gamma}x),
\end{equation}
where $\gamma$ is a scaling parameter and the dimensionless mass of the soliton is proportional to $\sqrt{\gamma}$, while the dimensionless radius is proportional to $1/\sqrt{\gamma}$. The dimensionless density $\vert\psi\vert^2$ and dimensionless radius $x$ can be transformed into dimensionful quantities by
\begin{align}
    \rho &= \mathcal{M}\mathcal{L}^{-3}\vert\psi\vert^2, \label{eq:density_conv} \\
    r &= \mathcal{L}x, \label{eq:mass_conv}
\end{align}
where
\begin{equation}\label{eq:length}
    \mathcal{L}=\left(\frac{8\pi\hbar^2}{3 m^2H_0^2\Omega_{m_0}}\right)^{\frac{1}{4}}\approx121\left(\frac{10^{-23}\operatorname{eV}}{m}\right)^{\frac{1}{2}}\operatorname{kpc},
\end{equation}
and 
\begin{align}\label{eq:mass}
    \mathcal{M}&=\frac{1}{G}\left(\frac{8\pi}{3 H_0^2\Omega_{m_0}}\right)^{-\frac{1}{4}}\left(\frac{\hbar}{m}\right)^{\frac{3}{2}}\nonumber\\
    &\approx 7\times 10^7\left(\frac{10^{-23}\operatorname{eV}}{m}\right)^{\frac{3}{2}}\operatorname{M}_{\odot}.
\end{align}

Ref.~\cite{Robles:2018fur} gives a piecewise parameterization of the generic ULDM profile 
\begin{equation}\label{eq:piecewise}
     \rho(r)=
    \begin{cases}
      \rho_{\mathrm{sol}}(r), & 0\leq r \leq r_{\alpha} \\
      \rho_\mathrm{NFW}(r), & r_{\alpha}\leq r \leq r_{\mathrm{vir}},
    \end{cases}
\end{equation}
where $\rho_{\mathrm{sol}}(r)$ is the appropriately scaled density profile of the ground state soliton solution. The contribution from the solitonic core and the overall virial mass is predicted to obey a scaling relationship \cite{Schive:2014hza, Chavanis:2019faf} which sets the central density, $\rho_c$, of a ULDM halo with virial mass, $\mathrm{M_{vir}}$. This yields an expression relating the core size to the velocity dispersion, and finally to the halo virial mass.%
\footnote{The authors of \cite{Schive:2014hza} suggest the following general expression:
\begin{equation}
    M_c = \alpha \left(\vert E\vert/M\right)^{1/2},
\end{equation}
where the core mass $M_c$ is determined by the total energy, $E$, and the total mass of the halo, $M$ where $\alpha$ is a constant of order unity. They then explain that the right hand side of the equation represents the halo velocity dispersion, while the left hand side  represents the inverse core size due to soliton scaling laws. By invoking the virial condition of the spherical collapse model, the authors then  construct the redshift dependent relationship between the solitonic core mass and the halo virial mass for a ULDM halo.}

\begin{figure}[t]
\centering
\includegraphics[scale = 0.6, trim={1cm 0cm 0cm 0.4cm}]{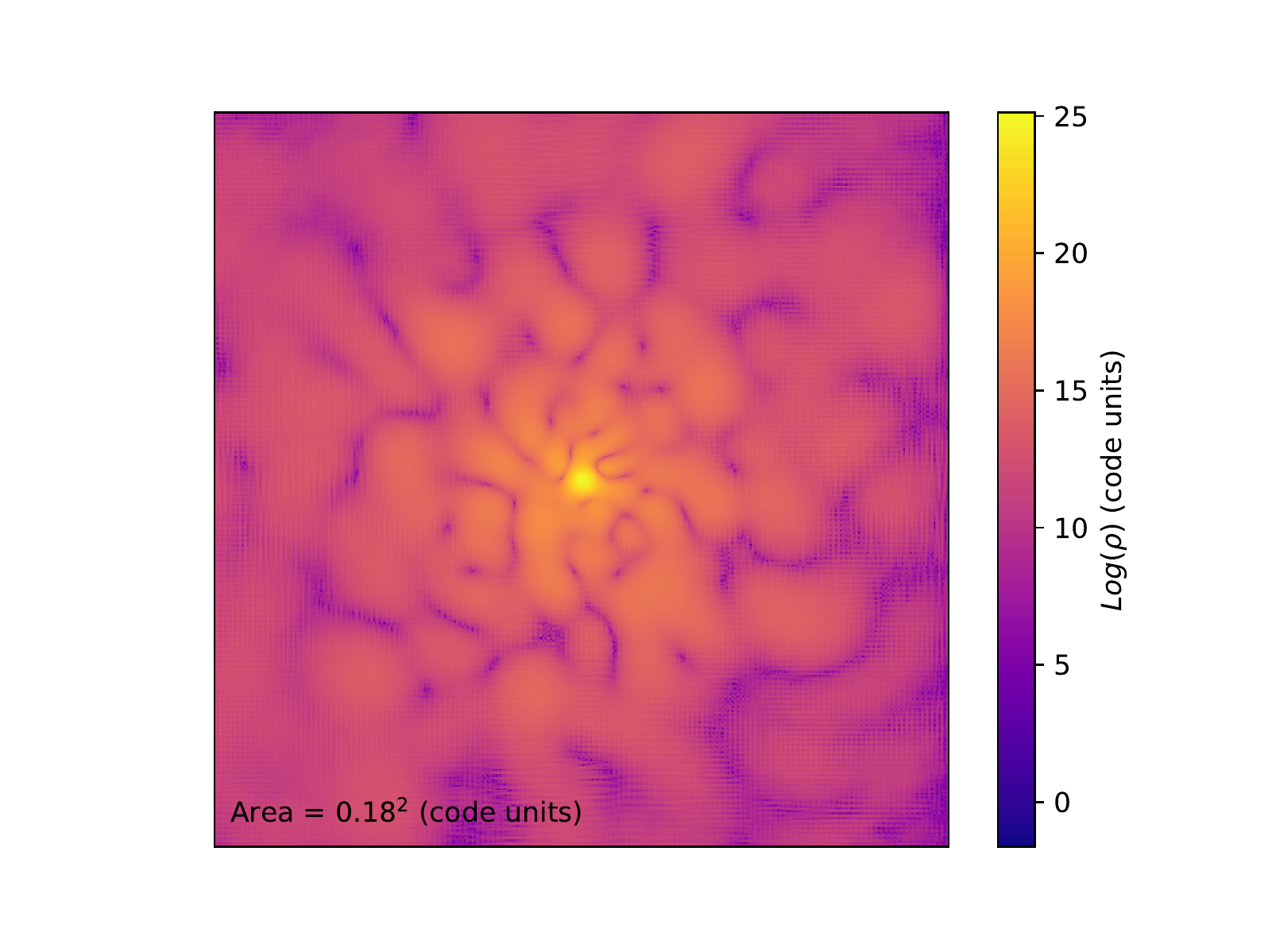}
\caption{Illustration of the scale of the fluctuations present in the incoherent outer halo for a merger of 8 randomly located solitons. The contour plot represents the ($\mathrm{log}_{10}$ scaled) local density across a slice through the centre of the final halo. In this plot, distance is not log-scaled, and we see that the spatial size of the fluctuations is of the same order of magnitude as the solitonic core itself.}\label{fig:contour}
\end{figure}
This core-halo mass relation can also be understood simply by matching the virial velocities of the core and the wider halo (see Appendix \ref{app:core-halo} for details). 
At $z=0$ the relationship is found to be \cite{Schive:2014hza} 
\begin{equation}\label{eq:central_dens}
    \rho_c = 2.94\times10^6 \operatorname{M}_{\odot}\operatorname{kpc}^{-3}\left(\frac{\mathrm{M_{vir}}}{10^9 \mathrm{M_{\odot}}}\right)^{4/3}\mathrm{m_{22}}^{2},
\end{equation}
and 
\begin{equation}
    r_c = 1.6 \operatorname{kpc}\left(\frac{\mathrm{M_{vir}}}{10^9 \mathrm{M_{\odot}}}\right)^{-1/3}\frac{1}{\mathrm{m_{22}}},
\end{equation}
where $r_c$ is the radius at which the density is half of the central value, and $\mathrm{m_{22}}$ is given by $\mathrm{m_{22}} \equiv m / 10^{-22} \operatorname{eV}$ where $m$ is the ULDM particle mass. 

While the piecewise semi-analytic ULDM profile is a useful tool, one should be mindful of its  limitations. For example, while a number of studies have attempted to establish  `universal' properties of ULDM halos, many of these analyses generated ULDM halos through the mergers of smaller compact objects  \cite{Schwabe:2016rze, Mocz:2017wlg}. This method of halo assembly is not representative of a realistic structure formation process, however it has the advantage of avoiding the computational difficulty of undertaking large-scale ULDM cosmological simulations. For this reason there is currently limited information from which to draw robust conclusions about the properties of astrophysical ULDM halos. In particular, universal application of the core-halo mass relation cannot be fully justified until more work is done to understand the characteristic timescales associated with the formation of quantum pressure-supported cores in scenarios including condensation from a fluctuating background, gravitational collapse in an expanding background, and mergers of objects with and without stable central cores. Moreover, it is difficult to accurately predict the effect that baryonic feedback will have on the formation of solitonic cores in halos of different masses, which could be significant at small radii in the present context.

Halo substructure is likewise missing from the semi-analytic model presented above. In simulations of soliton mergers the resulting halos have turbulent outer regions, with fluctuations on scales comparable to the core size, as illustrated in Figure \ref{fig:contour}. In addition to the fluctuations inherent in a large ULDM halo, smaller halos are likely to orbit or interact with larger halos. This   substructure is not captured by the semi-analytic model described above, and predictions for tracer velocity profiles may thus not match those of realistic astrophysical objects. Furthermore, temporal fluctuations in the core density are also missing from the semi-analytic model. Realistic halo cores are not exact soliton solutions of the Schr\"{o}dinger-Poisson equation,  they interact non-trivially with the fluctuating NFW-like outer halo, and their central densities can vary with time by as much as a factor of two \cite{Veltmaat:2018dfz}.

Taken together, these limitations suggest that the core-halo mass relation of the semi-analytic model should not be interpreted as an inviolable rule, but as a statement about the averaged characteristics of a statistical distribution. To estimate the variance corresponding to this distribution, we can consider a range of possible central densities for a given virial mass (somewhat analogous to the scatter in NFW concentration parameters \cite{Maccio:2008pcd}). The results of Ref.~\cite{Schive:2014hza}  indicate that a scatter in the core mass $\mathrm{M_c}$ of up to $\pm 50\%$ is possible for a given virial mass. Unfortunately, the small sample size and limited halo mass range ($ \mathrm{M_{vir}} \approx 10^8-10^{11} \operatorname{M}_{\odot}$) found in \cite{Schive:2014hza}  preclude a detailed analysis of the statistical properties of realistic astrophysical halos, but future simulations (especially those including baryonic feedback) should lead to improved predictions for this distribution. 

To partially account for statistical variance in halo properties, one can allow for variation in the radius at which the solitonic profile of the ULDM halo transitions into an NFW profile. This is acknowledged in Ref~\cite{Robles:2018fur} and is captured by the parameter $\alpha$: the transition radius, $r_{\alpha}$, is given by $r_{\alpha} = \alpha r_c$, with $3 \leq \alpha \leq 4$. For a given theoretical halo, an adjustment to the transition radius should be accompanied by changes in the parameters of the outer NFW halo, so as to maintain the core-halo mass ratio.

Thus, by taking the central soliton density and transition radius as variable parameters, we can create a range of plausible ULDM halo profiles for a given halo by using the virial mass to predict $\rho_c$, and assuming variation of $\pm 50\% $ around this central value. Given specific values for the central density and transition parameter $\alpha$, the solitonic piece of the ULDM profile is then completely specified, and its mass can be calculated. The remainder of the virial mass must be accounted for by the NFW tail of the profile. By matching the densities of the NFW tail to the inner soliton at the transition radius, the values of $R_s$ and $\rho_0$ for the NFW tail are obtained.  

%%%%%%%%%%%%%%%%%%%%%%%%%%% TO HERE

\section{ULDM and CDM halos and astrophysical data}\label{sec:velocity}

\begin{figure*}[t]
\centering
\includegraphics[scale=0.4, trim={2cm 0cm 0cm 1cm}]{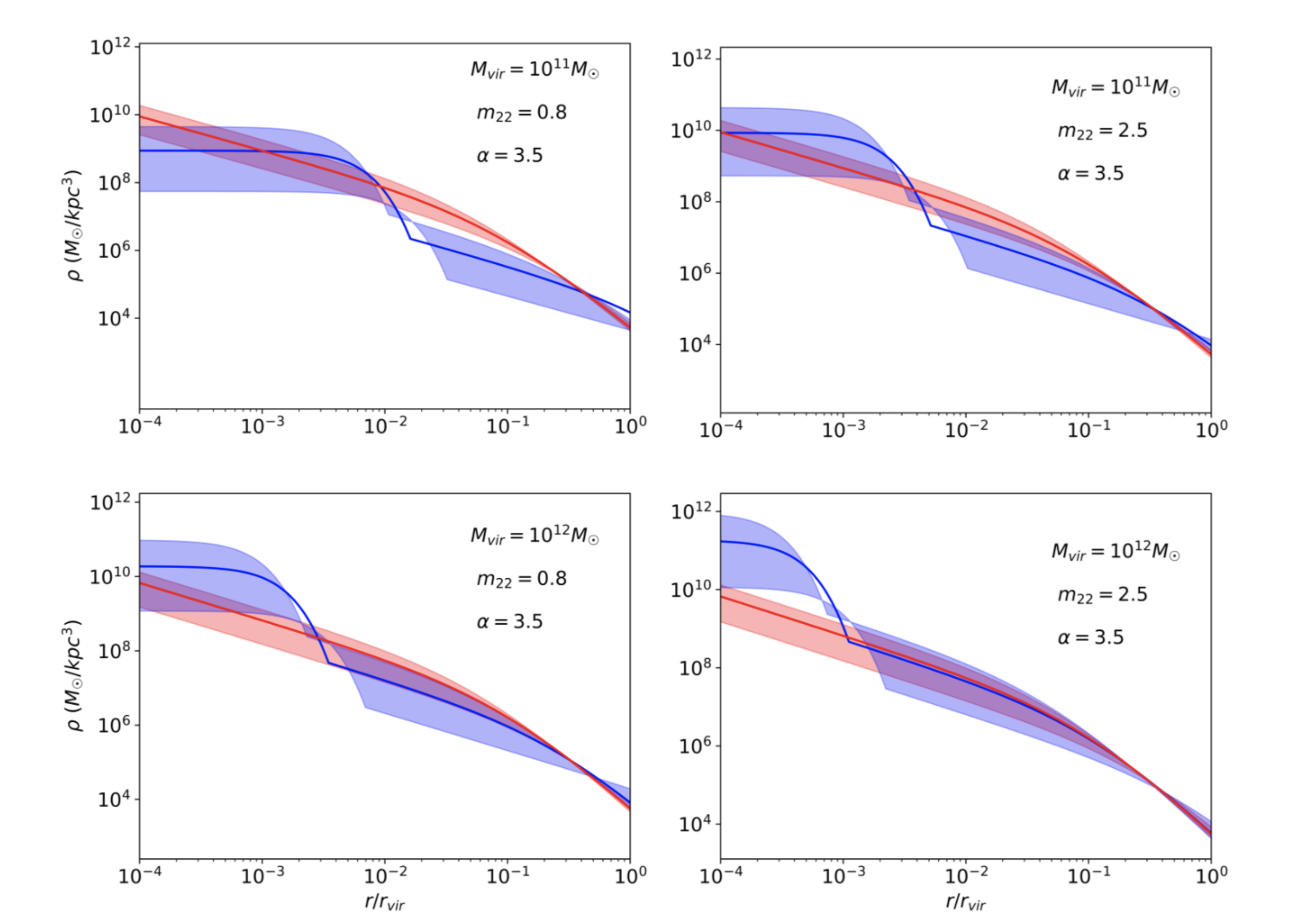}
\caption{Density profiles as a function of radius (normalised to the virial radius) for ULDM and NFW halos of masses $10^{11}\operatorname{M}_{\odot}$ (top) and $10^{12}\operatorname{M}_{\odot}$ (bottom). The left panel represents the results for $\mathrm{m_{22}} = 0.8$, while the right panel corresponds to $\mathrm{m_{22}}=2.5$. The transition radius is fixed at $r_{\alpha} = 3.5*r_c$. The blue shaded region represents the ULDM profile with $\operatorname{M}_c = \operatorname{M}_{\mathrm{cp}} \pm 50 \% \operatorname{M}_{\mathrm{cp}}$, while the solid blue line represents the ULDM profile when the theoretical core-halo mass relation is taken to be exact. The red shaded region represents the range of NFW profiles for a halo of the same virial mass with a 2$\sigma$ variation around the median (solid red line).}\label{fig:profiles}
\end{figure*}
 
We now compare the radial profiles of ULDM halos to NFW halos using the semi-analytic profiles described above, focusing on masses in the range $10^{11}$ and $10^{12} \operatorname{M}_{\odot}$, which may show an apparent worsening of the core-cusp problem \cite{Robles:2018fur}. Figure \ref{fig:profiles} compares such halos; the shaded blue region represents the ULDM halos for which the core-halo mass relation has a scatter $\operatorname{M}_c = \operatorname{M}_{\mathrm{cp}} \pm 50 \% $ range, where $\operatorname{M}_{\mathrm{cp}}$ is the theoretical prediction for the core mass. Note that because higher central densities correspond to narrower soliton profiles, the shaded region possesses `crossover points' near the transition from the solitonic to NFW profile, appearing somewhat skewed from the median line. Were we to vary more parameters in the model (such as transition radius and axion mass), we would see a broadening of the shaded region, such that the median line would be fully encompassed by the shaded region. Because we are here focusing primarily on the core mass (and therefore central density), we illustrate only the changes in density profile attributable to this, hence the restricted range of profiles shown as the shaded blue region.

The  Schr{\"o}dinger-Poisson soliton scaling relations show that the $\operatorname{M}_c = \operatorname{M}_{\mathrm{cp}} \pm 50 \% $ mass range corresponds to a range of $ \gamma_p /4 \leq \gamma \leq 9\gamma_p/4$, where $\gamma_p$ is the theoretical prediction of the square root of the dimensionless central density. Consequently, there is a large variation in the central density and thus widely varying predictions for the ULDM profiles. We fix $\alpha = 3.5$ (in the middle of the predicted range) which does not affect the central density as the core lies well within the solitonic region. Changing the value of $\alpha$ will, however, affect the predicted velocity profiles for each halo. We do not attempt to fit this parameter to data in this Section; the previously discussed limitations of the semi-analytic models employed here suggest that this would be unlikely to be a meaningful exercise. The blue ULDM profiles are compared to the red shaded regions of Figure \ref{fig:profiles}, showing the $2\sigma$ variation about the theoretical prediction for the concentration parameter of the NFW halo with the same virial mass \cite{Maccio:2008pcd}.

Following Ref~\cite{Robles:2018fur}, we plot to a minimum radius of $r/r_{\mathrm{vir}} = 10^{-4}$ and for the same choices of $\mathrm{m_{22}}$. For any $\operatorname{M}_{\mathrm{vir}}$, the NFW halo density  will inevitably exceed that of the ULDM halo at very small radii, though the threshold for this transition may be arbitrarily small, and not observationally relevant. However, we note that the apparent worsening of the core-cusp discrepancy does depend on the choice of inner radial cutoff.

From Figure \ref{fig:profiles} we see that for halo masses of $10^{11}\operatorname{M}_{\odot}$ there is a wide range of $\mathrm{M_c}$ for which the ULDM profile is `less cuspy' than its NFW counterpart. For a halo mass of $10^{12}\operatorname{M}_{\odot}$ and a ULDM particle mass $\mathrm{m_{22}}=0.8$ the range of plausible ULDM profiles likewise includes those which are `less cuspy' than the corresponding NFW profile. At higher particle mass ($\mathrm{m_{22}}=2.5$) for $10^{12}\operatorname{M}_{\odot}$ halos, the NFW profiles tend to be less peaked than corresponding ULDM profiles at radial distances in  the range $10^{-4}\leq r/r_{\mathrm{vir}} \leq 1$.  

To assess the suitability of these semi-analytic profiles, we compare to observations drawn from the SPARC database. Because observations yield the (line of sight) velocity distributions of tracer stars as a function of galactocentric radius rather than the halo density itself, we must first transform our theoretical density profiles into velocity profiles. In so doing, we acknowledge that the effects of non-circular motion and kinematic irregularities constitute a non-trivial source of random error in observed velocities, which should be kept in mind especially when working with limited data sets.

\begin{figure*}[t]
\centering
\includegraphics[scale=0.9, trim={0cm 2.5cm 3cm 0cm}]{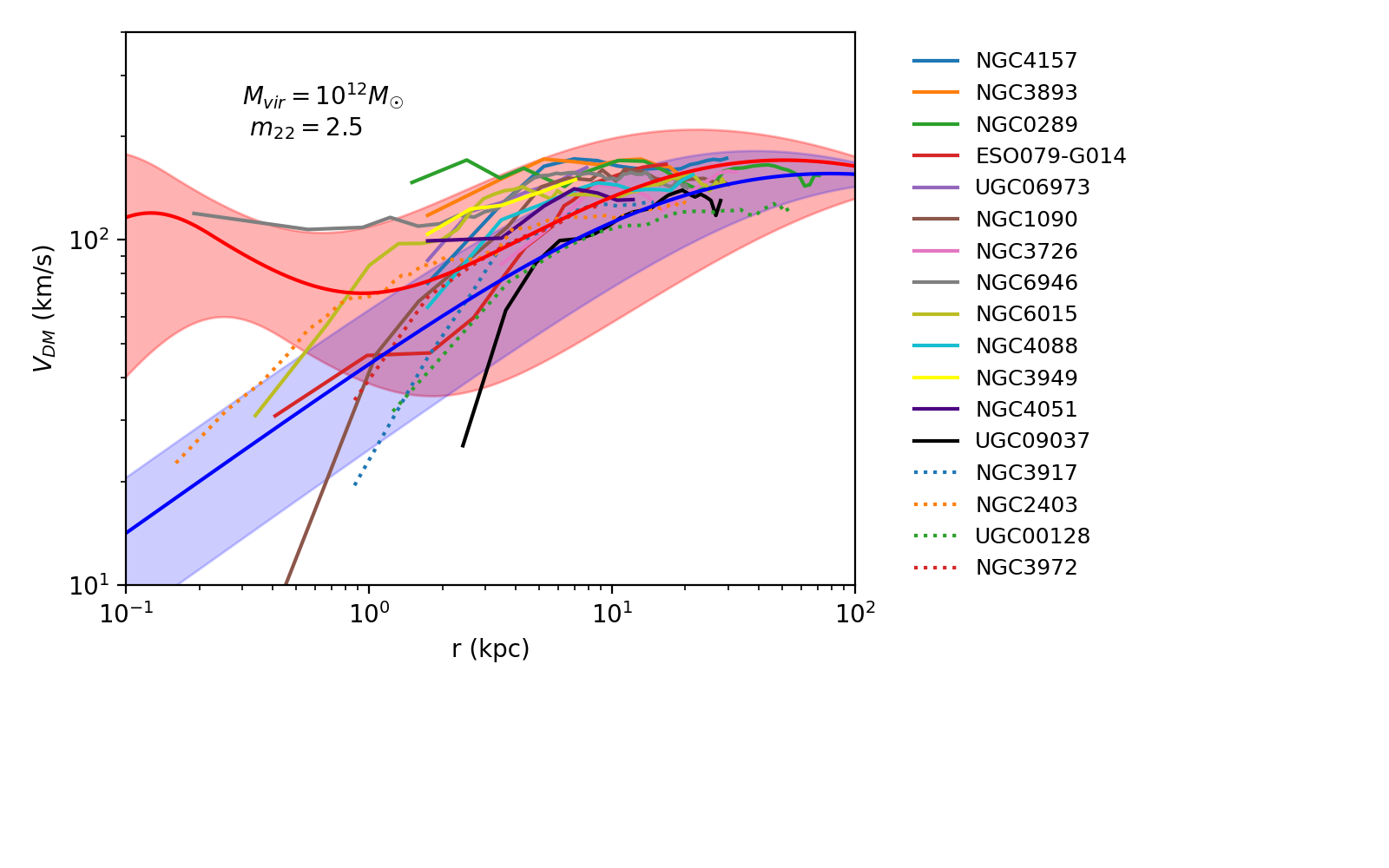}
\caption{Velocity distributions for galaxies with maximum velocities in the range $125 \leq v < 175\operatorname{kms}^{-1}$ in the SPARC database. Data at innermost radii is limited for these galaxies, making it difficult to determine the overall characteristics of the profiles. The SPARC data is plotted alongside theoretical NFW (shaded blue) and ULDM (shaded red) profiles, assuming a virial mass of $10^{12} \mathrm{M}_{\odot}$, $\mathrm{m_{22}} = 2.5$, and $\pm 50 \%$ scatter in the ULDM core-halo mass relation and $\pm2\sigma$ scatter in NFW concentration. Galaxies in the legend are ordered from highest maximum velocity (top) to lowest (bottom).}\label{fig:high_v} 
\end{figure*}

We convert  density profiles to velocity distributions \cite{Sofue:2008wt} via 
\begin{equation}
    V(r)^2 = \frac{4\pi G}{r}\int_0^r \rho(r')r'^2 dr',
\end{equation}
where 
\begin{equation}\label{eq:vel_decomp}
    V^2 = V_{\mathrm{disk}}^2 + V_{\mathrm{bulge}}^2 + V_{\mathrm{gas}}^2 + V_{\mathrm{halo}}^2.
\end{equation}
The SPARC database contains photometric data for 175 galaxies and rotation curves from $\mathrm{H}_{\mathrm{I}}$/$\mathrm{H}_{\alpha}$ studies. The disk and bulge velocities in the SPARC database are given for $\Upsilon = 1 \operatorname{M}_{\odot}/\operatorname{L}_{\odot}$ at $3.6\operatorname{\mu m}$. However, the greatest source of uncertainty in mass modelling is the assumed stellar mass-to-light ratio, $\Upsilon_\star$ \cite{Lelli:2016zqa}. As in \cite{Robles:2018fur}, we  assume a constant value of $\Upsilon_\star = 0.2 \operatorname{M}_{\odot}/\operatorname{L}_{\odot}$ at $3.6\operatorname{\mu m}$, likewise noting that  this constitutes a non-trivial source of uncertainty. Moreover, there is significant uncertainty in the SPARC data itself. Error bars are omitted in the following graphs for ease of viewing, however, they are discussed in Appendix\ref{sec:errors}. 

The characteristics of the velocity profiles in the SPARC database vary widely from galaxy to galaxy; however we qualitatively identify two subsets of galaxies; those with maximum tracer velocities $75 \leq v < 125\operatorname{kms}^{-1}$, and those for which $125 \leq v < 175\operatorname{kms}^{-1}$. The former group tends to exhibit a strong steepening in the radial velocity profile toward the inner halo, while the  profiles for the latter group are comparatively flat\footnote{We exclude data for which the velocities calculated according to Equation \ref{eq:vel_decomp} are inconsistent - this can occur due to the uncertainty in the assumption for $\Upsilon_\star$.}. We assume that higher asymptotic velocities correspond to a larger halo mass, and consider halo masses in the range $10^{11} - 10^{12} \operatorname{M}_{\odot}$, expecting that masses at the top end of the range will give a better match to galaxies with higher asymptotic velocities.

In Figure \ref{fig:high_v}, we see that galaxies with asymptotic velocities at the higher end of the range do not always exhibit a pronounced steepening of the velocity profile at small radii. Indeed in some cases there is simply no data at small radii. From this figure we see that while a halo mass of $10^{12} \mathrm{M}_{\odot}$ with $\mathrm{m_{22}} = 2.5$ provides a reasonable fit to the data at radii $>10 \mathrm{kpc}$, it is difficult to judge the fit at small radii, where the ULDM and NFW profiles differ most strongly, due to a lack of data. Furthermore, while the data at higher radii seems to be relatively clustered, there are significant deviations within the limited data that exists at small radii. For example, the curves for NGC1090 and NGC6946 are widely disparate at small radii, but seem to converge at larger radii. Attempting to fit such data to a single set of model parameters would be of limited utility without a much more comprehensive data set from which statistical outliers could be properly identified. Furthermore, we note that there are substantial changes in theoretical ULDM velocity profiles under variation in the ULDM particle mass. The scale of these changes is exhibited in \ref{app:mass-vel}. As such, we remark that analyses of the sort presented here would benefit greatly from tighter constraints on the ULDM particle mass. 

\begin{figure*}[t]
\centering
\includegraphics[scale=0.9, trim={0.5cm 1cm 0cm 1cm}]{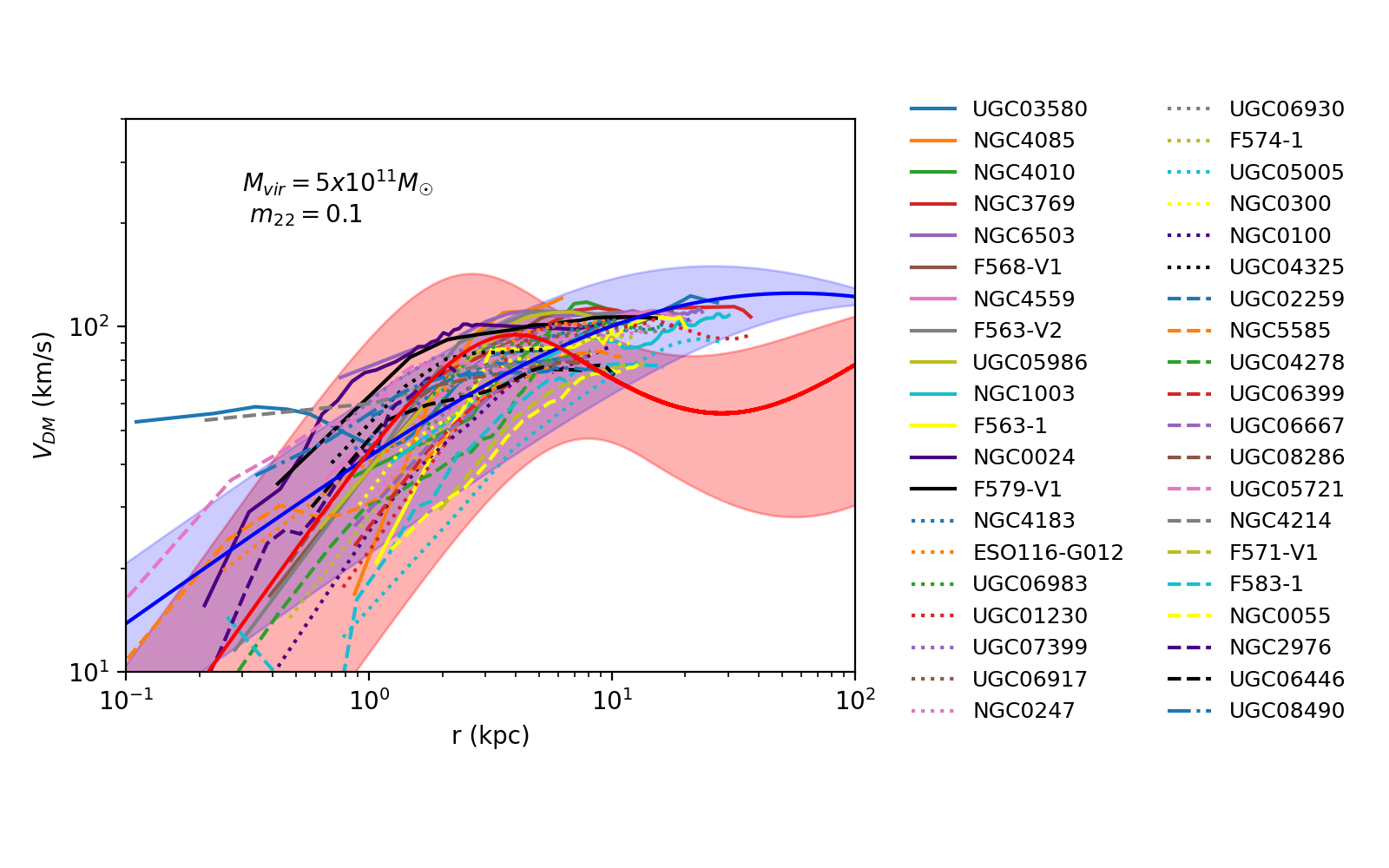}
\caption{Velocity distributions for galaxies with maximum velocities in the range $75 \leq v < 125\operatorname{kms}^{-1}$ in the SPARC database. Data at outer radii is limited for these galaxies, making it difficult to determine the overall characteristics of the profiles. The SPARC data is plotted along with theoretical NFW (shaded blue) and ULDM (shaded red) profiles, assuming a virial mass of $5\times10^{11} \mathrm{M}_{\odot}$, $\mathrm{m_{22}} = 0.1$, and $\pm 50 \%$ scatter in the ULDM core-halo mass relation and $\pm2\sigma$ scatter in NFW concentration. Galaxies in the legend are ordered from highest maximum velocity (top left) to lowest (bottom right).}\label{fig:low_v}
\end{figure*}

By contrast, for galaxies with smaller maximum velocities ($75 \leq v < 125\operatorname{kms}^{-1}$), there is more data at smaller radii. For such galaxies we see the steepening rotation curves characteristic of cored density profiles, as shown in Figure \ref{fig:low_v}. In this case, choosing parameters such that the theoretical profiles overlap with the data at small radii is easy (in this case $\mathrm{m_{22}} = 0.1$, $\mathrm{M}_{\mathrm{vir}}5\times10^{11} \mathrm{M}_{\odot}$), however it is not clear whether the behaviour of this profile would fit data at larger radii were it available. Furthermore, while the choice $\mathrm{m_{22}} = 0.1$ provides a reasonable fit to the data in this case, a ULDM particle mass $\mathrm{m_{22}} = 0.1$ is in tension with constraints from the Lyman-$\alpha$ forest, as well as high-redshift UV luminosity function comparisons\footnote{$ 0.8 < \mathrm{m_{22}} < 2.5$ is preferred by current constraints, as mentioned in \cite{Robles:2018fur}} \cite{Amendola:2005ad, Bozek:2014uqa, Armengaud:2017nkf, Ni:2019qfa, Nebrin:2018vqt}. It must be acknowledged, however, that baryonic feedback is expected to have the greatest impact in the innermost regions of realistic halos. As such, agreement between our semi-analytic DM-only model and observational data at small radii should be interpreted cautiously, especially since this is also the region where assumptions regarding the stellar mass to light ratio have the greatest significance.

\section{Conclusions}\label{sec:conclusion}

The ULDM model has gained attention in part because it may offer a solution to the CDM core-cusp problem. However, in some cases ULDM profiles can actually have higher densities than their NFW counterparts at observationally relevant radii in the interior of halos with mass $\mathrm{M} \gtrsim 10^{12} \mathrm{M}_{\odot}$, where the central density is determined by the theoretical core-halo mass relation. However, apparent statistical spread in the ULDM core-halo mass relation \cite{Schive:2014hza} leads to a sizeable range of plausible central densities for a halo of any given mass. Furthermore, analyses of oscillations of the cores of ULDM halos on timescales much smaller than the relaxation time have demonstrated significant fluctuations in central density \cite{Veltmaat:2018dfz}. This suggests that theoretical core-halo mass relations should not be interpreted too literally for any individual ULDM halo, and this should be taken into account when performing model-selection analyses. The limited available simulation data means that the exact features of the distribution of halo properties in ULDM are poorly characterised. Nevertheless, it remains apparent that core masses at the lower end of the plausible range could mitigate the apparent worsening of the core-cusp discrepancy for ULDM halos.  

When the spread in the theoretical core-halo relation is accounted for, comparisons of theoretical ULDM and NFW profiles to the SPARC database yield inconclusive results as far as interior regions of the halos are concerned. Parameters can be easily chosen to provide a superficial fit to given subsets of data. However, the available data often do not span a large enough range of radial values to assess the relative merits of the UDLM and NFW profiles over the whole profile. In particular, rotation curves in the SPARC database which exhibit a strong steepening at small radii often lack data at large radii, whereas rotation curves which include large radii information often lack measurements at small radii. From our restricted analysis, however, it appears that neither the theoretical ULDM nor CDM model can reliably reproduce the data across a broad range. Wide deviations at small galactocentric radii cannot be accounted for by random measurement errors, and suggest that sophisticated modelling of baryonic physics will be necessary before any conclusions can be reasonably drawn. The lack of baryonic physics in both the semi-analytic CDM and ULDM models is significant, as are other limitations due to poorly characterised statistics and simplistic assumptions about halo modelling. 

In principle, one could perform a BIC analysis to determine which of the ULDM or NFW models is more strongly favoured by the data \cite{Liddle:2004nh}. The model with the lowest BIC, defined as
\begin{equation}
    \mathrm{BIC} = k \ln{\mathrm{N}} - 2 \ln{\mathcal{L}},    
\end{equation}
is preferred. Here $\mathcal{L}$, $\mathrm{N}$, and $k$ are the maximised likelihood function, the sample size, and the number of model parameters, respectively. The utility of BIC or other model selection tools, however, is hampered by the lack of comprehensive data, the high number of free parameters (the stellar mass to light ratio in the SPARC data, assumed virial mass of the galactic halos,  ULDM particle mass, the NFW concentration parameter, the UDLM soliton to NFW transition radius and  variation in the ULDM core-halo mass relation) and the omission of baryonic feedback in ULDM simulations. Indeed, BIC analyses are known to be compromised when the sample size of the data is not sufficiently large in comparison to the number of free parameters in the model. Because of this limitation, and large and unquantified systematic biases in both the observational data and theoretical predictions, such analyses are premature at this point. Previous studies such as Ref.~\cite{Bar2018acw} of the core-halo mass relation and the fitting of semi-analytical profiles to galaxy data also emphasise the necessarily preliminary and tentative nature of all analyses of ULDM-derived rotation curves. 

To summarise, the parameter space describing ``typical'' ULDM halos is larger than simple semi-analytical models suggest. It is necessary to constrain this parameter space in order to make robust model selection possible. Tightening the constraints on the plausible ULDM particle mass \cite{Castellano:2019hdd, Lidz:2018fqo, Davoudiasl:2019nlo} and obtaining additional spectroscopic data with improved uncertainties covering a greater halo mass range and radius would be of tremendous benefit in this regard. Such improved data can be expected from future surveys \cite{Simon:2019kmm}. In addition, better ULDM cosmological structure formation simulations are needed to  improve the understanding of ULDM halo evolution \cite{Lin:2018whl, Clough:2018exo, Mocz:2015sda} and these should also incorporate baryonic feedback. Thus, the key conclusion to be drawn from this work is that more information from simulations and astrophysical observations is needed, as is more sophisticated incorporation of baryonic effects within semi-analytic models of both ULDM and CDM, before the relative successes of each model can be fairly compared.

\begin{acknowledgements}
We acknowledge invaluable discussions with Jens Niemeyer, Shaun Hotchkiss, and Mateja Gosenca in completing this work. We also acknowledge support from the Marsden Fund of the Royal Society of New Zealand. This research was supported by use of the Nectar Research Cloud, a collaborative Australian research platform supported by the National Collaborative Research Infrastructure Strategy (NCRIS).

\end{acknowledgements}

\begin{appendix}

\begin{figure*}[t]
\centering
\includegraphics[scale = 0.8, trim={0cm 2.5cm 1cm 0cm}]{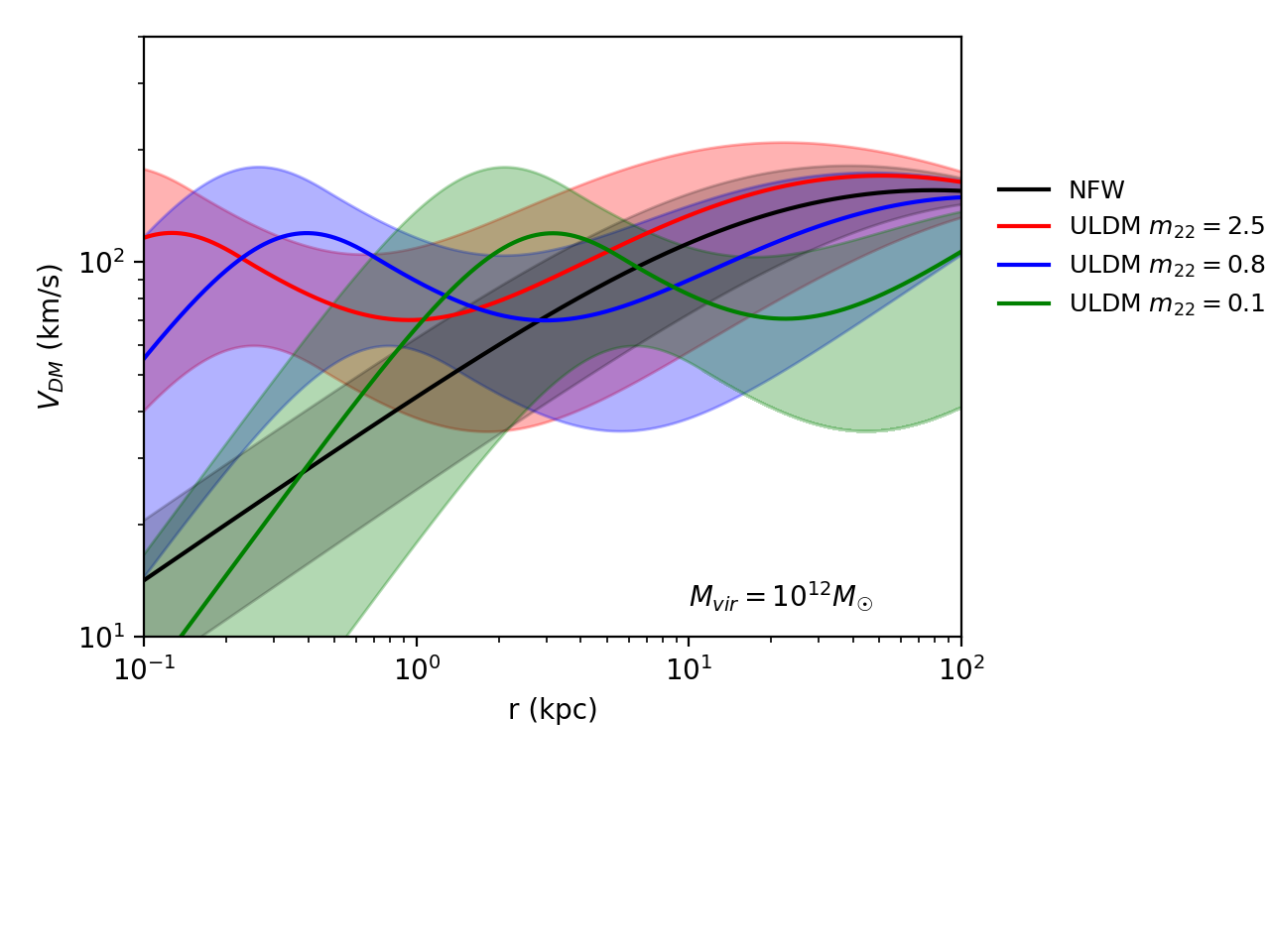} 
\caption{Plot demonstrating the effect of changing the ULDM particle mass assumption on the velocity profiles for halos of mass $10^{12}\mathrm{M}_{\odot}$.}\label{fig:vel_5_10_11}
\end{figure*}

\section{Core-halo mass relation}\label{app:core-halo}

The core-halo mass relation can be simply interpreted as the statement that the average internal velocity of a tracer mass in the core must be equal to the virial velocity of a tracer mass in the wider halo. If this were not the case, and instead the average velocity were higher within the core, these higher velocity particles would move outward, resulting in dynamical mass redistribution within the halo. During this process, the halo would not be in equilibrium and would thus not be virialised.

From the virial theorem we have that $E_K=-1/2 \ E_P$, where $E_K$ and $E_P$ represent kinetic and potential energies, respectively. Alternatively we can write:

\begin{equation}
    \frac{1}{2}M_{tot}v^2=\frac{1}{4}\frac{GM_{tot}^2}{R_{tot}},
\end{equation}
where $G$ is the gravitational constant, $M_{tot}$ and $R_{tot}$ are the total mass and radius, and $v^2$ is the mean of the squares of individual tracer velocities. Demanding that $v^2$ is the same for the core as for the total virialised halo allows us to then write
\begin{align}\label{eq:virial_cond}
    v^2&=\frac{G\mathrm{M_{vir}}}{2 R_{\mathrm{vir}}}=\frac{G M_{core}}{2 R_{core}}\nonumber\\
    &\rightarrow R_{core}=\frac{M_{core} R_{\mathrm{vir}}}{\mathrm{M_{vir}}}.
\end{align}
We know from the soliton scaling properties that $R_{core}\propto M_{core}^{-1}$, and since $\mathrm{M_{vir}}=4/3 \ \pi R_{\mathrm{vir}}^3 \Bar{\rho}$, we also have $R_{\mathrm{vir}} \propto \mathrm{M_{vir}}^{1/3}$. Hence, Equation \ref{eq:virial_cond} becomes
\begin{align}
    &R_{core}^2\propto \frac{R_{\mathrm{vir}}}{\mathrm{M_{vir}}}\nonumber\\
    &\rightarrow R_{core}^2\propto \frac{\mathrm{M_{vir}}^{1/3}}{\mathrm{M_{vir}}}\nonumber\\
    &\rightarrow R_{core}\propto\left(\mathrm{M_{vir}}^{-2/3}\right)^{1/2}\nonumber\\
    &\rightarrow R_{core}\propto \mathrm{M_{vir}}^{-1/3}.
\end{align}
With this scaling relation in mind, the constant of proportionality may be determined through analysis of simulated halos.

\section{ - Impact of ULDM particle mass on halo velocity profiles}\label{app:mass-vel}

Figure \ref{fig:vel_5_10_11} demonstrates the scale of the changes to the velocity profiles of theoretical ULDM halos under changes in the ULDM particle mass. In order to perform a meaningful parameter fitting exercise, observational data would be required to span the wide range of radii illustrated here. In this way, the regions of the rotation curves most sensitive to the assumption for the ULDM particle mass could be tested simultaneously. Presently available data, when binned according to e.g. maximum velocity, is likely to yield disparate preferences for the ULDM particle mass, as illustrated in Figures \ref{fig:high_v} and \ref{fig:low_v}. Further work to constrain the plausible range of the particle mass will make comparisons of the ULDM and CDM models with astrophysical data more effective.

\section{Errors in SPARC data}\label{sec:errors}

As discussed in Section \ref{sec:velocity}, the uncertainties associated with the SPARC rotation curves for the galaxies studied here make it difficult to draw robust conclusions as to the suitability of one or the other dark matter model. Indeed, sources of error quoted in the SPARC database relate not only to the random error in measured line-of-site velocities, but also to errors on the galaxy distance measurement, inclination, and total luminosity. Furthermore, inaccuracy in the assumed stellar mass-to-light ratio may lead to skewed velocity decompositions, a systematic effect that could exceed the stochastic measurement errors.

\begin{figure*}[t]
\centering
\includegraphics[scale=0.9, trim={0.5cm 3cm 0cm 1cm}]{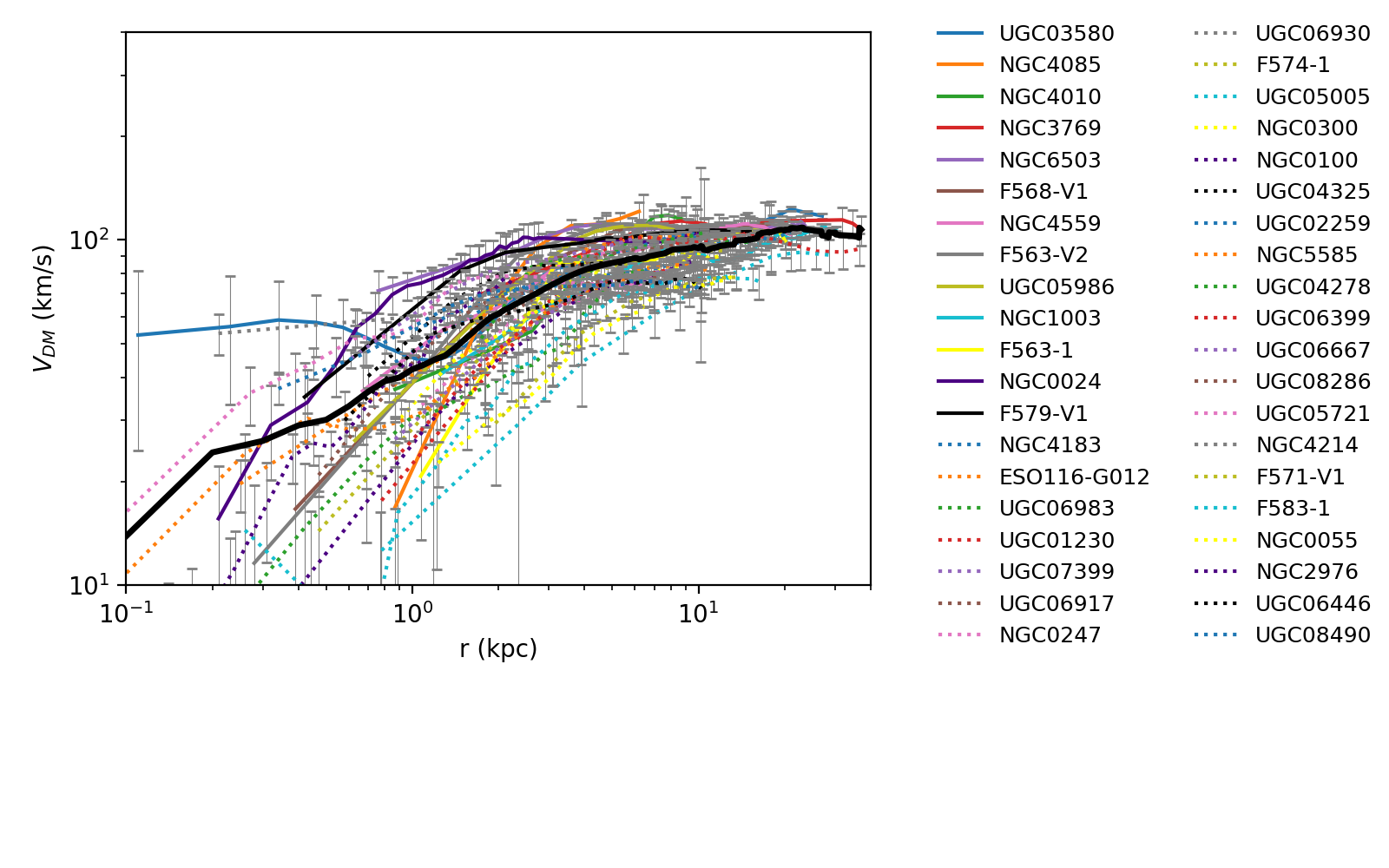}
\caption{Radial distributions for galaxies with maximum velocities in the range $75 \leq v < 125\operatorname{kms}^{-1}$ in the SPARC database. The average velocity curve is shown by the bold black line. Large uncertainties coupled with a wide spread of data at small radii limit the utility of this data set for determining the precise details of small scale structure in dark matter halos.}\label{fig:low_v_err}
\end{figure*}

\begin{figure*}[t]
\centering
\includegraphics[scale=0.9, trim={0.5cm 3cm 0cm 1cm}]{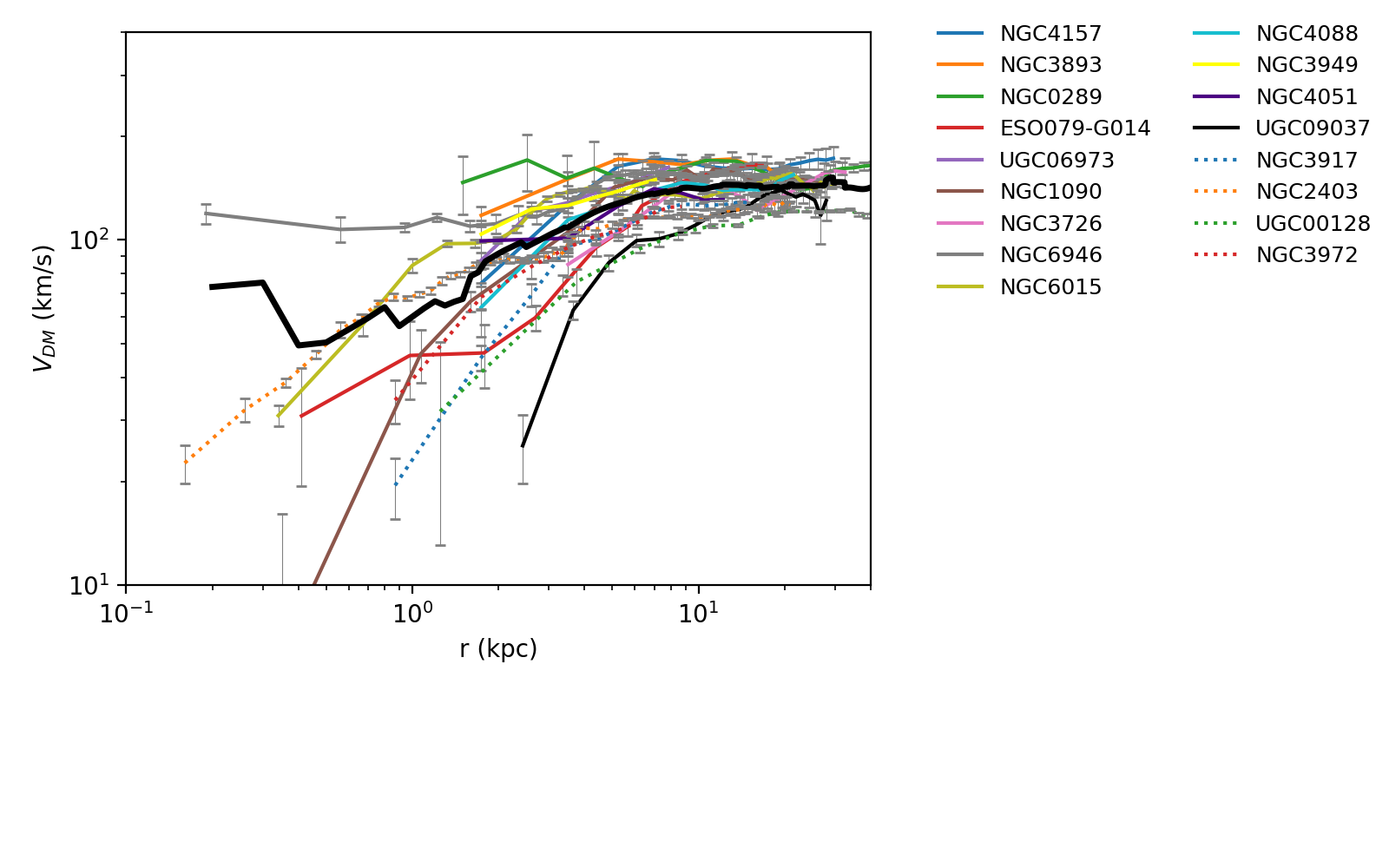}
\caption{Radial distributions for galaxies with maximum velocities in the range $125 \leq v < 175\operatorname{kms}^{-1}$ in the SPARC database. The average profile is shown in the bold black line. The limited number of galaxies with high asymptotic velocities makes it difficult to judge typical galaxy characteristics in this regime. Furthermore, we see that data at small radii is lacking, and for that which is available, the variation in velocity profiles cannot be accounted for by random error in the measured velocities alone.}\label{fig:high_v_err}
\end{figure*}

Figures \ref{fig:low_v_err} and \ref{fig:high_v_err} show the error bars associated with the low asymptotic velocity ($75 \leq v < 125\operatorname{kms}^{-1}$) and high asymptotic velocity ($125 \leq v < 175\operatorname{kms}^{-1}$) measurements, respectively. In Figure \ref{fig:low_v_err}, we see large error bars at small radii. It is precisely this regime in which accurate velocity profiles are needed to assess the significance of the core-cusp discrepancy - a key differentiating factor between ULDM and CDM models as illustrated in Figure \ref{fig:low_v}. Hence, more comprehensive, accurate data in this regime would be of tremendous benefit. Furthermore, in Figure \ref{fig:low_v} we also observe that at higher radii, the ULDM model exhibits a characteristic dip in the radial velocity profile. The observational data does not extend far enough into the high radius regime to reveal whether such features are exist in astrophysical structures. Indeed, the data in this case tapers off at galactocentric radii exceeding around 10kpc. From the limited number of galaxies for which data approaches this regime (in particular  UGC03580, UGC01230, NGC3769, NGC1003), there does not seem to be a tendency toward a dip. The error bars are relatively constrained in this high radius region, as shown in Figure \ref{fig:low_v_err}, so the absence of a dip arguably weakens support for the ULDM model. More data at higher radii is required to make a strong determination on this point. 

Meanwhile, in Figure \ref{fig:high_v_err}, the spread of data at high radius is within the scale of the error bars, but at smaller radii the data is not encompassed by random measurement error alone. This may suggest that grouping galaxies by asymptotic velocities alone is insufficient as a method of characterisation. This spread of data may suggest that grouping galaxies based on asymptotic velocities alone is an insufficient method of characterisation. However, there is very limited data in this sample at small radii, so both larger data sets and comprehensive modelling of baryonic effects in high density inner regions are required to resolve this issue.

\end{appendix}

\bibstyle{pasa-mnras}
\bibliography{1r_lamboo_notes}

\begin{thebibliography}{48}
\providecommand{\natexlab}[1]{#1}
\providecommand{\url}[1]{\texttt{#1}}
\expandafter\ifx\csname urlstyle\endcsname\relax
  \providecommand{\doi}[1]{doi: #1}\else
  \providecommand{\doi}{doi: \begingroup \urlstyle{rm}\Url}\fi

\bibitem[Ade et~al.(2016)]{Ade:2015xua}
P.~A.~R. Ade et~al.
\newblock {Planck 2015 results. XIII. Cosmological parameters}.
\newblock \emph{Astron. Astrophys.}, 594:\penalty0 A13, 2016.
\newblock \doi{10.1051/0004-6361/201525830}.

\bibitem[Amendola and Barbieri(2006)]{Amendola:2005ad}
L.~Amendola and R.~Barbieri.
\newblock {Dark matter from an ultra-light pseudo-Goldsone-boson}.
\newblock \emph{Phys. Lett.}, B642:\penalty0 192--196, 2006.
\newblock \doi{10.1016/j.physletb.2006.08.069}.

\bibitem[Armengaud et~al.(2017)Armengaud, Palanque-Delabrouille, Yèche, Marsh,
  and Baur]{Armengaud:2017nkf}
E.~Armengaud, N.~Palanque-Delabrouille, C.~Yèche, D.~J.~E. Marsh, and J.~Baur.
\newblock {Constraining the mass of light bosonic dark matter using SDSS
  Lyman-$\alpha$ forest}.
\newblock \emph{Mon. Not. Roy. Astron. Soc.}, 471\penalty0 (4):\penalty0
  4606--4614, 2017.
\newblock \doi{10.1093/mnras/stx1870}.

\bibitem[Bar et~al.(2018)Bar, Blas, Blum, and Sibiryakov]{Bar2018acw}
N.~Bar, D.~Blas, K.~Blum, and S.~Sibiryakov.
\newblock {Galactic rotation curves versus ultralight dark matter: Implications
  of the soliton-host halo relation}.
\newblock \emph{Phys. Rev.}, D98\penalty0 (8):\penalty0 083027, 2018.
\newblock \doi{10.1103/PhysRevD.98.083027}.

\bibitem[{Benitez-Llambay} et~al.(2018){Benitez-Llambay}, {Frenk}, {Ludlow},
  and {Navarro}]{Benitez-Llambay:2018}
A.~{Benitez-Llambay}, C.~S. {Frenk}, A.~D. {Ludlow}, and J.~F. {Navarro}.
\newblock {Baryon-induced dark matter cores in the EAGLE simulations}.
\newblock \emph{arXiv e-prints}, art. arXiv:1810.04186, Oct 2018.

\bibitem[Bozek et~al.(2015)Bozek, Marsh, Silk, and Wyse]{Bozek:2014uqa}
B.~Bozek, D.~J.~E. Marsh, J.~Silk, and R.~F.~G. Wyse.
\newblock {Galaxy UV-luminosity function and reionization constraints on axion
  dark matter}.
\newblock \emph{Mon. Not. Roy. Astron. Soc.}, 450\penalty0 (1):\penalty0
  209--222, 2015.
\newblock \doi{10.1093/mnras/stv624}.

\bibitem[Castellano et~al.(2018)Castellano, Menci, Grazian, Merle, Sanchez,
  Schneider, and Totzauer]{Castellano:2019hdd}
M.~Castellano, N.~Menci, A.~Grazian, A.~Merle, N.~G. Sanchez, A.~Schneider, and
  M.~Totzauer.
\newblock {Constraining Dark Matter models with extremely distant galaxies}.
\newblock In \emph{{Proceedings, Vulcano Workshop 2018: Frontier Objects in
  Astrophysics and Particle Physics: Vulcano Island, Sicily, Italy, May 20-26,
  2018}}, pages 200--213, 2018.

\bibitem[Chavanis(2019)]{Chavanis:2019faf}
P.-H. Chavanis.
\newblock {Derivation of the core mass -- halo mass relation of fermionic and
  bosonic dark matter halos from an effective thermodynamical model}.
\newblock 2019.

\bibitem[Clough et~al.(2018)Clough, Dietrich, and Niemeyer]{Clough:2018exo}
K.~Clough, T.~Dietrich, and J.~C. Niemeyer.
\newblock {Axion star collisions with black holes and neutron stars in full 3D
  numerical relativity}.
\newblock \emph{Phys. Rev.}, D98\penalty0 (8):\penalty0 083020, 2018.
\newblock \doi{10.1103/PhysRevD.98.083020}.

\bibitem[{Dashyan} et~al.(2018){Dashyan}, {Silk}, {Mamon}, {Dubois}, and
  {Hartwig}]{2018MNRAS.473.5698D}
G.~{Dashyan}, J.~{Silk}, G.~A. {Mamon}, Y.~{Dubois}, and T.~{Hartwig}.
\newblock {AGN feedback in dwarf galaxies?}
\newblock \emph{"Mon. Not. Roy. Astron. Soc."}, 473\penalty0 (4):\penalty0
  5698--5703, Feb 2018.
\newblock \doi{10.1093/mnras/stx2716}.

\bibitem[Davoudiasl and Denton(2019)]{Davoudiasl:2019nlo}
H.~Davoudiasl and P.~B. Denton.
\newblock {Ultra Light Boson Dark Matter and Event Horizon Telescope
  Observations of M87*}.
\newblock 2019.

\bibitem[de~Bernardis et~al.(2000)]{deBernardis:2000sbo}
P.~de~Bernardis et~al.
\newblock {A Flat universe from high resolution maps of the cosmic microwave
  background radiation}.
\newblock \emph{Nature}, 404:\penalty0 955--959, 2000.
\newblock \doi{10.1038/35010035}.

\bibitem[Dutton et~al.(2018)Dutton, Macciò, Buck, Dixon, Blank, and
  Obreja]{Dutton:2018nop}
A.~A. Dutton, A.~V. Macciò, T.~Buck, K.~L. Dixon, M.~Blank, and A.~Obreja.
\newblock {NIHAO XX: The impact of the star formation threshold on the
  cusp-core transformation of cold dark matter haloes}.
\newblock 2018.

\bibitem[{Genina} et~al.(2018){Genina}, {Ben{\'\i}tez-Llambay}, {Frenk},
  {Cole}, {Fattahi}, {Navarro}, {Oman}, {Sawala}, and {Theuns}]{Genina:2018}
A.~{Genina}, A.~{Ben{\'\i}tez-Llambay}, C.~S. {Frenk}, S.~{Cole}, A.~{Fattahi},
  J.~F. {Navarro}, K.~A. {Oman}, T.~{Sawala}, and T.~{Theuns}.
\newblock {The core-cusp problem: a matter of perspective}.
\newblock \emph{"Mon. Not. Roy. Astron. Soc."}, 474:\penalty0 1398--1411, Feb
  2018.
\newblock \doi{10.1093/mnras/stx2855}.

\bibitem[Halverson et~al.(2002)]{Halverson:2001yy}
N.~W. Halverson et~al.
\newblock {DASI first results: A Measurement of the cosmic microwave background
  angular power spectrum}.
\newblock \emph{Astrophys. J.}, 568:\penalty0 38--45, 2002.
\newblock \doi{10.1086/338879}.

\bibitem[Hanany et~al.(2000)]{Hanany:2000qf}
S.~Hanany et~al.
\newblock {MAXIMA-1: A Measurement of the cosmic microwave background
  anisotropy on angular scales of 10 arcminutes to 5 degrees}.
\newblock \emph{Astrophys. J.}, 545:\penalty0 L5, 2000.
\newblock \doi{10.1086/317322}.

\bibitem[Herrera et~al.(2017)Herrera, Waga, and Jorás]{Herrera:2017epn}
D.~Herrera, I.~Waga, and S.~E. Jorás.
\newblock {Calculation of the critical overdensity in the spherical-collapse
  approximation}.
\newblock \emph{Phys. Rev.}, D95\penalty0 (6):\penalty0 064029, 2017.
\newblock \doi{10.1103/PhysRevD.95.064029}.

\bibitem[Hu and Dodelson(2002)]{Hu:2001bc}
W.~Hu and S.~Dodelson.
\newblock {Cosmic microwave background anisotropies}.
\newblock \emph{Ann. Rev. Astron. Astrophys.}, 40:\penalty0 171--216, 2002.
\newblock \doi{10.1146/annurev.astro.40.060401.093926}.

\bibitem[Hui et~al.(2017)Hui, Ostriker, Tremaine, and Witten]{Hui:2016ltb}
L.~Hui, J.~P. Ostriker, S.~Tremaine, and E.~Witten.
\newblock {Ultralight scalars as cosmological dark matter}.
\newblock \emph{Phys. Rev.}, D95\penalty0 (4):\penalty0 043541, 2017.
\newblock \doi{10.1103/PhysRevD.95.043541}.

\bibitem[Lee et~al.(2001)]{Lee:2001yp}
A.~T. Lee et~al.
\newblock {A High spatial resolution analysis of the MAXIMA-1 cosmic microwave
  background anisotropy data}.
\newblock \emph{Astrophys. J.}, 561:\penalty0 L1--L6, 2001.
\newblock \doi{10.1086/324437}.

\bibitem[Lelli et~al.(2016)Lelli, McGaugh, and Schombert]{Lelli:2016zqa}
F.~Lelli, S.~S. McGaugh, and J.~M. Schombert.
\newblock {SPARC: Mass Models for 175 Disk Galaxies with Spitzer Photometry and
  Accurate Rotation Curves}.
\newblock \emph{Astron. J.}, 152:\penalty0 157, 2016.
\newblock \doi{10.3847/0004-6256/152/6/157}.

\bibitem[Liddle(2004)]{Liddle:2004nh}
A.~R. Liddle.
\newblock {How many cosmological parameters?}
\newblock \emph{Mon. Not. Roy. Astron. Soc.}, 351:\penalty0 L49--L53, 2004.
\newblock \doi{10.1111/j.1365-2966.2004.08033.x}.

\bibitem[Lidz and Hui(2018)]{Lidz:2018fqo}
A.~Lidz and L.~Hui.
\newblock {Implications of a prereionization 21-cm absorption signal for fuzzy
  dark matter}.
\newblock \emph{Phys. Rev.}, D98\penalty0 (2):\penalty0 023011, 2018.
\newblock \doi{10.1103/PhysRevD.98.023011}.

\bibitem[Lin et~al.(2018)Lin, Schive, Wong, and Chiueh]{Lin:2018whl}
S.-C. Lin, H.-Y. Schive, S.-K. Wong, and T.~Chiueh.
\newblock {Self-consistent construction of virialized wave dark matter halos}.
\newblock \emph{Phys. Rev.}, D97\penalty0 (10):\penalty0 103523, 2018.
\newblock \doi{10.1103/PhysRevD.97.103523}.

\bibitem[Ludlow et~al.(2014)Ludlow, Navarro, Angulo, Boylan-Kolchin, Springel,
  Frenk, and White]{Ludlow:2013vxa}
A.~D. Ludlow, J.~F. Navarro, R.~E. Angulo, M.~Boylan-Kolchin, V.~Springel,
  C.~Frenk, and S.~D.~M. White.
\newblock {The mass–concentration–redshift relation of cold dark matter
  haloes}.
\newblock \emph{Mon. Not. Roy. Astron. Soc.}, 441\penalty0 (1):\penalty0
  378--388, 2014.
\newblock \doi{10.1093/mnras/stu483}.

\bibitem[Maccio' et~al.(2008)Maccio', Dutton, and Bosch]{Maccio:2008pcd}
A.~V. Maccio', A.~A. Dutton, and F.~C. v.~d. Bosch.
\newblock {Concentration, Spin and Shape of Dark Matter Haloes as a Function of
  the Cosmological Model: WMAP1, WMAP3 and WMAP5 results}.
\newblock \emph{Mon. Not. Roy. Astron. Soc.}, 391:\penalty0 1940--1954, 2008.
\newblock \doi{10.1111/j.1365-2966.2008.14029.x}.

\bibitem[Mocz and Succi(2015)]{Mocz:2015sda}
P.~Mocz and S.~Succi.
\newblock {Numerical solution of the nonlinear Schrödinger equation using
  smoothed-particle hydrodynamics}.
\newblock \emph{Phys. Rev.}, E91\penalty0 (5):\penalty0 053304, 2015.
\newblock \doi{10.1103/PhysRevE.91.053304}.

\bibitem[Mocz et~al.(2017)Mocz, Vogelsberger, Robles, Zavala, Boylan-Kolchin,
  Fialkov, and Hernquist]{Mocz:2017wlg}
P.~Mocz, M.~Vogelsberger, V.~H. Robles, J.~Zavala, M.~Boylan-Kolchin,
  A.~Fialkov, and L.~Hernquist.
\newblock {Galaxy formation with BECDM – I. Turbulence and relaxation of
  idealized haloes}.
\newblock \emph{Mon. Not. Roy. Astron. Soc.}, 471\penalty0 (4):\penalty0
  4559--4570, 2017.
\newblock \doi{10.1093/mnras/stx1887}.

\bibitem[Moore(1994)]{Moore:1994yx}
B.~Moore.
\newblock {Evidence against dissipationless dark matter from observations of
  galaxy haloes}.
\newblock \emph{Nature}, 370:\penalty0 629, 1994.
\newblock \doi{10.1038/370629a0}.

\bibitem[Navarro et~al.(1996)Navarro, Frenk, and White]{Navarro:1995iw}
J.~F. Navarro, C.~S. Frenk, and S.~D.~M. White.
\newblock {The Structure of cold dark matter halos}.
\newblock \emph{Astrophys. J.}, 462:\penalty0 563--575, 1996.
\newblock \doi{10.1086/177173}.

\bibitem[Nebrin et~al.(2020)Nebrin, Ghara, and Mellema]{Nebrin:2018vqt}
O.~Nebrin, R.~Ghara, and G.~Mellema.
\newblock {Fuzzy Dark Matter at Cosmic Dawn: New 21-cm Constraints}.
\newblock \emph{JCAP}, 2019\penalty0 (04):\penalty0 051, 2020.
\newblock \doi{10.1088/1475-7516/2019/04/051}.

\bibitem[Netterfield et~al.(2002)]{Netterfield:2001yq}
C.~B. Netterfield et~al.
\newblock {A measurement by Boomerang of multiple peaks in the angular power
  spectrum of the cosmic microwave background}.
\newblock \emph{Astrophys. J.}, 571:\penalty0 604--614, 2002.
\newblock \doi{10.1086/340118}.

\bibitem[Ni et~al.(2019)Ni, Wang, Feng, and Di~Matteo]{Ni:2019qfa}
Y.~Ni, M.-Y. Wang, Y.~Feng, and T.~Di~Matteo.
\newblock {Predictions for the Abundance of High-redshift Galaxies in a Fuzzy
  Dark Matter Universe}.
\newblock 2019.

\bibitem[Ragagnin et~al.(2018)Ragagnin, Dolag, Moscardini, Biviano, and
  D'Onofrio]{Ragagnin:2018enf}
A.~Ragagnin, K.~Dolag, L.~Moscardini, A.~Biviano, and M.~D'Onofrio.
\newblock {Dependency of halo concentration on mass, redshift and fossilness in
  Magneticum hydrodynamic simulations}.
\newblock 2018.
\newblock \doi{10.1093/mnras/stz1103}.

\bibitem[Read et~al.(2018)Read, Walker, and Steger]{Read:2018pft}
J.~I. Read, M.~G. Walker, and P.~Steger.
\newblock {The case for a cold dark matter cusp in Draco}.
\newblock \emph{Mon. Not. Roy. Astron. Soc.}, 481\penalty0 (1):\penalty0
  860--877, 2018.
\newblock \doi{10.1093/mnras/sty2286}.

\bibitem[{Richings} et~al.(2018){Richings}, {Frenk}, {Jenkins}, and
  {Robertson}]{Richings:2018}
J.~{Richings}, C.~{Frenk}, A.~{Jenkins}, and A.~{Robertson}.
\newblock {Subhalo destruction in the Apostle and Auriga simulations}.
\newblock \emph{arXiv e-prints}, art. arXiv:1811.12437, Nov 2018.

\bibitem[Robles et~al.(2019)Robles, Bullock, and
  Boylan-Kolchin]{Robles:2018fur}
V.~H. Robles, J.~S. Bullock, and M.~Boylan-Kolchin.
\newblock {Scalar Field Dark Matter: Helping or Hurting Small-Scale Problems in
  Cosmology?}
\newblock \emph{Mon. Not. Roy. Astron. Soc.}, 483\penalty0 (1):\penalty0
  289--298, 2019.
\newblock \doi{10.1093/mnras/sty3190}.

\bibitem[Schive et~al.(2014)Schive, Liao, Woo, Wong, Chiueh, Broadhurst, and
  Hwang]{Schive:2014hza}
H.-Y. Schive, M.-H. Liao, T.-P. Woo, S.-K. Wong, T.~Chiueh, T.~Broadhurst, and
  W.~Y.~P. Hwang.
\newblock {Understanding the Core-Halo Relation of Quantum Wave Dark Matter
  from 3D Simulations}.
\newblock \emph{Phys. Rev. Lett.}, 113\penalty0 (26):\penalty0 261302, 2014.
\newblock \doi{10.1103/PhysRevLett.113.261302}.

\bibitem[Schumann(2019)]{Schumann:2019eaa}
M.~Schumann.
\newblock {Direct Detection of WIMP Dark Matter: Concepts and Status}.
\newblock 2019.

\bibitem[Schwabe et~al.(2016)Schwabe, Niemeyer, and Engels]{Schwabe:2016rze}
B.~Schwabe, J.~C. Niemeyer, and J.~F. Engels.
\newblock {Simulations of solitonic core mergers in ultralight axion dark
  matter cosmologies}.
\newblock \emph{Phys. Rev.}, D94\penalty0 (4):\penalty0 043513, 2016.
\newblock \doi{10.1103/PhysRevD.94.043513}.

\bibitem[Simon et~al.(2019)]{Simon:2019kmm}
J.~D. Simon et~al.
\newblock {Testing the Nature of Dark Matter with Extremely Large Telescopes}.
\newblock 2019.

\bibitem[Sofue et~al.(2009)Sofue, Honma, and Omodaka]{Sofue:2008wt}
Y.~Sofue, M.~Honma, and T.~Omodaka.
\newblock {Unified Rotation Curve of the Galaxy -- Decomposition into de
  Vaucouleurs Bulge, Disk, Dark Halo, and the 9-kpc Rotation Dip --}.
\newblock \emph{Publ. Astron. Soc. Jap.}, 61:\penalty0 227, 2009.
\newblock \doi{10.1093/pasj/61.2.227}.

\bibitem[Springel et~al.(2005)]{Springel:2005nw}
V.~Springel et~al.
\newblock Simulating the joint evolution of quasars, galaxies and their
  large-scale distribution.
\newblock \emph{Nature}, 435:\penalty0 629--636, 2005.
\newblock \doi{10.1038/nature03597}.

\bibitem[Suto et~al.(2016)Suto, Kitayama, Osato, Sasaki, and
  Suto]{Suto:2015jdt}
D.~Suto, T.~Kitayama, K.~Osato, S.~Sasaki, and Y.~Suto.
\newblock {Confrontation of Top-Hat Spherical Collapse against Dark Halos from
  Cosmological N-Body Simulations}.
\newblock \emph{Publ. Astron. Soc. Jap.}, 68\penalty0 (1):\penalty0 14, 2016.
\newblock \doi{10.1093/pasj/psv122}.

\bibitem[Veltmaat et~al.(2018)Veltmaat, Niemeyer, and
  Schwabe]{Veltmaat:2018dfz}
J.~Veltmaat, J.~C. Niemeyer, and B.~Schwabe.
\newblock {Formation and structure of ultralight bosonic dark matter halos}.
\newblock \emph{Phys. Rev.}, D98\penalty0 (4):\penalty0 043509, 2018.
\newblock \doi{10.1103/PhysRevD.98.043509}.

\bibitem[{Watkins} et~al.(2019){Watkins}, {van der Marel}, {Sohn}, and
  {Evans}]{Watkins2019ApJ}
L.~L. {Watkins}, R.~P. {van der Marel}, S.~T. {Sohn}, and N.~W. {Evans}.
\newblock {Evidence for an Intermediate-mass Milky Way from Gaia DR2 Halo
  Globular Cluster Motions}.
\newblock \emph{\apj}, 873\penalty0 (2):\penalty0 118, Mar 2019.
\newblock \doi{10.3847/1538-4357/ab089f}.

\bibitem[Weinberg et~al.(2015)Weinberg, Bullock, Governato, Kuzio~de Naray, and
  Peter]{Weinberg:2013aya}
D.~H. Weinberg, J.~S. Bullock, F.~Governato, R.~Kuzio~de Naray, and A.~H.~G.
  Peter.
\newblock {Cold dark matter: controversies on small scales}.
\newblock \emph{Proc. Nat. Acad. Sci.}, 112:\penalty0 12249--12255, 2015.
\newblock \doi{10.1073/pnas.1308716112}.

\bibitem[White(2001)]{White:2000jv}
M.~J. White.
\newblock {The Mass of a halo}.
\newblock \emph{Astron. Astrophys.}, 367:\penalty0 27, 2001.
\newblock \doi{10.1051/0004-6361:20000357}.

\end{thebibliography}

\end{document}